\documentclass[english,10pt,twocolumn, aps,prd,floatfix,nofootinbib,superscriptaddress, notitlepage]{revtex4-2} 

\usepackage[usenames,dvipsnames]{color}  
\usepackage{graphicx}
\usepackage{setspace}
\usepackage{caption}
\usepackage{amsmath}
\usepackage{slashed}
\usepackage{amssymb}
\usepackage[colorlinks=true,citecolor=darkred,urlcolor=darkred, pdfborder={0 0 0}]{hyperref}
\usepackage[normalem]{ulem}
\usepackage{xcolor}
\usepackage{float}
\usepackage{microtype}
\usepackage{lipsum}

\makeatletter
\def\p@subsection{}
\makeatother

\definecolor{darkred}{rgb}{0.6,0,0}

\definecolor{linkcolor}{rgb}{0,0,0.5}

\usepackage[T1]{fontenc} 
\usepackage{subcaption}
\usepackage{float}

\usepackage[compat=1.1.0]{tikz-feynhand}
\usepackage{tikz-feynman}
\tikzfeynmanset{compat=1.1.0}
\usepackage{feynmp}
\usepackage{tikzsymbols}
\usepackage{array}
\usepackage{pifont} 
\usepackage{amsmath}
\usepackage{mathrsfs}
\usepackage{placeins}

\definecolor{linkcolor}{rgb}{0,0,0.5}

\begin{document}

\title{\textbf{Neutrino Mass Sum Rules from Modular $\mathcal{A}_4$ Symmetry}}
\author{Salvador Centelles Chuliá}
\affiliation{Max-Planck-Institut f\"ur Kernphysik, Saupfercheckweg 1, 69117 Heidelberg, Germany}
\author{Ranjeet Kumar}
\affiliation{Department of Physics, Indian Institute of Science Education and Research - Bhopal \\
Bhopal ByPass Road, Bhauri, Bhopal 462066, India}
\author{Oleg Popov}
\affiliation{Department of Biology, Shenzhen MSU-BIT University,\\ 1, International University Park Road, Shenzhen 518172, China}
\author{Rahul Srivastava}
\affiliation{Department of Physics, Indian Institute of Science Education and Research - Bhopal \\
Bhopal ByPass Road, Bhauri, Bhopal 462066, India}


\begin{abstract}
  \vspace{1cm} 
Modular symmetries offer a dynamic approach to understanding the flavour structure of leptonic mixing. Using the modular $\mathcal{A}_4$ flavour symmetry integrated in a type-II seesaw, we propose a simple and minimalistic model that restricts the neutrino oscillation parameter space and, most importantly, introduces a sum rule in the physical neutrino masses. When combined with the mass squared differences observed in neutrino oscillations, this sum rule determines the absolute neutrino mass scale. This has significant implications for cosmology, neutrinoless double beta decay experiments and direct neutrino mass measurements.  In particular, the model predicts $\sum_i m_i \approx 0.1$ eV for both normal and inverted ordering, and thus can be fully probed by the current generation of cosmological probes in the upcoming years.

\end{abstract}
\maketitle

\section{Introduction}
\label{sec:intro}

More than two decades after the discovery of neutrino oscillations, the origin of their masses and mixing parameters is still not well understood. This challenge, known as the "leptonic flavour puzzle", stands as a central topic in both theoretical and experimental research. There are now several theoretical approaches and proposals in the literature, aimed at understanding the leptonic mass and mixing pattern. One of the most popular approach is to use new symmetries, often taken as non-abelian discrete groups, called "flavour symmetries", to obtain a deeper understanding of the leptonic flavour puzzle. The flavour symmetries often lead to predictions for the neutrino mixing angles and phases, which can be then tested in various experiments, particularly in neutrino oscillations. 

Another interesting consequence of flavour symmetries is occurrences of neutrino mass "sum rules". In their usual form these are relations between the complex eigenvalues of the neutrino mass matrix, which in turn can be read as a constraint on the physical neutrino masses and the Majorana phases, thus restricting, for example, the neutrinoless double beta decay ($0\nu ee$) parameter space. This approach enables a complementary path to neutrino oscillation experiments as a way to distinguish between given flavour models, as detailed in \cite{Barry:2010yk,King:2013psa,Gehrlein:2016wlc} and the references within. 

A further development which has recently gained considerable attention is to promote the flavour symmetry to a modular symmetry, first achieved in~\cite{Feruglio:2017spp}. In the modular symmetry approach all the Yukawa couplings are promoted to modular forms that transform non-trivially under the modular group. This approach leads to the prediction of the lepton and/or quark mixing angles. Several studies, since 2017's initial paper, then have been performed. They include $\mathcal{A}_4$ modular symmetry based works~\cite{deAnda:2018ecu,Nomura:2019yft,Ding:2019zxk,Okada:2019mjf,Nomura:2019lnr,Dasgupta:2021ggp,Zhang:2019ngf,Nomura:2019xsb,Kobayashi:2019gtp,Wang:2019xbo,Abbas:2020qzc,Okada:2020dmb,Okada:2020rjb,Behera:2020sfe,Asaka:2020tmo,Nagao:2020snm,Hutauruk:2020xtk,Yao:2020qyy,Chen:2021zty,Mishra:2022egy,Devi:2023vpe,Mishra:2023ekx}, $\mathcal{S}_4$ modular symmetry based  works~\cite{Penedo:2018nmg,Novichkov:2018ovf,Novichkov:2019sqv,Kobayashi:2019mna,Okada:2019lzv,Kobayashi:2019xvz,Wang:2019ovr,Wang:2020dbp,Zhao:2021jxg}, $\mathcal{A}_5$ modular symmetry based works~\cite{Novichkov:2018nkm,Ding:2019xna,Behera:2022wco,Behera:2021eut}, application of modular symmetries to quark and lepton sectors~\cite{Lu:2019vgm,Okada:2020rjb,Yao:2020qyy,Baur:2022hma,Feruglio:2023uof}, modular symmetries in the context of theories of the grand unification~\cite{deAnda:2018ecu,Zhao:2021jxg,Chen:2021zty}, modular symmetry based Dirac neutrino mass models~\cite{Dasgupta:2021ggp,Wang:2020dbp}, 
(generalised) CP symmetry and modular invariance~\cite{Baur:2019iai,Baur:2019kwi,Novichkov:2019sqv,Kobayashi:2019uyt,Meloni:2023aru,Behera:2022wco}, 
non-fine-tuned explanations of fermion mass hierarchies in modular-invariant models~\cite{Lu:2019vgm,King:2020qaj,Okada:2020ukr,Feruglio:2021dte,Novichkov:2021evw,Li:2023dvm}, 
modulus stabilisation~\cite{Novichkov:2022wvg}, 
$\mathcal{S}_3$ modular symmetry based~\cite{Mishra:2020gxg,Okada:2019xqk,Meloni:2023aru,Mishra:2023cjc},
eclectic flavour symmetry based~\cite{Baur:2022hma,Li:2023dvm},
leptonic sum rules with modular symmetries~\cite{Gehrlein:2020jnr},
W-mass anomaly related modular symmetries~\cite{Mishra:2023cjc},
modular symmetry models based on extra dimensional extensions and compactifications of the modular space~\cite{Kikuchi:2022txy},
and the strong CP problem related to modular invariance~\cite{Feruglio:2023uof}.

 In this work we aim to use the powerful framework of the modular symmetries to obtain a stronger version of the traditional sum rules, where a relation is obtained for the singular values of the mass matrix, i.e. the physical neutrino masses, instead of the complex eigenvalues as in \cite{Gehrlein:2020jnr}. Once the measured mass squared differences constraints are applied, the absolute neutrino mass scale is fixed. We derive this sum rule in a particular UV complete model in which the neutrino masses come from a type-II seesaw~\cite{Schechter:1980gr,Magg:1980ut,Cheng:1980qt,Mohapatra:1980yp} with the Yukawas transforming as modular forms under $\mathcal{A}_4$ modular symmetry. The model is simple and elegant with a $SU(2)_L$ triplet $\Delta$ and modulus $\tau$  being the only superfields added beyond the Minimal Supersymmetric Standard Model (MSSM). Yet our model is highly predictive where along with the sum rules we also obtain predictions for the leptonic mixing angles and CP phases and analyze their implications for various running and upcoming experiments.

This letter is organized as follows. In Sec.~\ref{sec:model} we present the field content of the model and their gauge and modular transformation rules. In Sec.~\ref{sec:mee} we show the consequences of the strongest prediction of the model, the neutrino mass sum rule, in  neutrinoless double beta decay experiments, cosmology and KATRIN. In Sec.~\ref{sec:oscillations} we flesh out the predictions of the model in neutrino oscillation experiments for both normal and inverted ordering. Finally we conclude in Sec.~\ref{sec:conclusions}.

\section{Model}
\label{sec:model}
In this section we discuss the model framework based on the finite modular group $\Gamma_3 \simeq \mathcal{A}_4 $. Neutrino masses and mixing will be generated from a type-II seesaw mechanism. Within our model framework, we assign the leptonic superfields $L_i$ and $E_i^c$ to be triplet  $\mathbf{(3)}$ and  singlets  $\mathbf{(1, 1', 1'')}$ under $\mathcal{A}_4$ respectively, with weights of -3 and -1. The Higgs doublets $H_u$ and $H_d$ are trivial singlets  $\mathbf{(1)}$ of $\mathcal{A}_4$ with weight 0. To accommodate the type-II seesaw mechanism we include a $SU(2)_L$ triplet $\Delta$, which is also a trivial singlet $\mathbf{(1)}$ of $\mathcal{A}_4$ with weight 0. The charge assignment of the superfields and their weights are summarized in Tab.~\ref{tab:particles}. 
The detailed discussion of modular symmetry and its transformation laws are given in App.~\ref{sec:modular}. A field $\phi^{(I)}$ transforms under the modular transformation of Eq.~\eqref{eq:field-SL2Z},  as
\begin{equation}
\label{eq:field-SL2Z}
\phi^{(I)} \to (c\tau+d)^{-\mathit{k}_I}\rho^{(I)}(\gamma)\phi^{(I)},
\end{equation}
where  $-\mathit{k}_I$ represents the modular weight and $\rho^{(I)}(\gamma)$ signifies an unitary representation matrix of $\gamma\in\Gamma(2)$.
\begin{table}[h]
    \centering
    \begin{tabular}{cccccc}
        \hline \hline
         Fields & $SU(3)_c$ & $SU(2)_L$ & $U(1)_Y$ & $\Gamma_3 \simeq \mathcal{A}_4$ & $-\mathit{k}$ \\ \hline
         $L_i$ & $1$ & $2$ & $-\frac{1}{2}$ & $\pmb{3}$ & $-3$ \\
         ${E^c_i}$ & $1$ & $1$ & $1$ & $\pmb{1, 1', 1''}$ & $-1$ \\ \hline
         $H_u$ & $1$ & $2$ & $\frac{1}{2}$ & $\pmb{1}$ & $0$ \\
         $H_d$ & $1$ & $2$ & $-\frac{1}{2}$ & $\pmb{1}$ & $0$  \\
         $\Delta$ & $1$ & $3$ & $1$ & $\pmb{1}$ & $0$\\
         \hline \hline
    \end{tabular}
    \caption{The charge assignments of the superfields under $SU(3)_C  \otimes SU(2)_L \otimes  U(1)_Y \otimes   \mathcal{A}_4$,  where $-\mathit{k}$ is modular weight.}
    \label{tab:particles}
\end{table}
\begin{table}[h]
    \centering
    \begin{tabular}{ccc}
        \hline \hline
         Yukawas & $\Gamma_3 \simeq \mathcal{A}_4$ &  $\pmb{\mathit{k}}$\\ \hline
        $\pmb{Y_e=Y_{{3}}^{(4)}}$ & $\pmb{3}$ & $4$ \\
        $\pmb{Y_{\nu, 1}=Y_{{3a}}^{(6)}}$ & $\pmb{3}$ & $6$ \\
        $\pmb{Y_{\nu,2}=Y_{{3b}}^{(6)}
        }$ & $\pmb{3}$ & $6$ \\
        \hline \hline
    \end{tabular}
    \caption{Modular transformations of Yukawa couplings and their weights.}
    \label{tab:Yukawa}
\end{table}

Under these symmetries the superpotential of our model is given as follows:
\begin{eqnarray}
 \label{eq:main_W}
\mathcal{W} & = & \alpha_1 \left( \pmb{Y_e}  L \right)_1 E^c_1 H_d + \alpha_2 \left( \pmb{Y_e}  L \right)_{1''} E^c_2 H_d \nonumber \\
& + &  \alpha_3 \left( \pmb{Y_e}  L \right)_{1'} E^c_3 H_d 
 +  \alpha \left( \pmb{Y_{\nu, 1}}  \left( L  L \right)_{3_S} \right)_1 \Delta \\
& + & \beta \left( \pmb{Y_{\nu, 2}}  \left( L  L \right)_{3_S} \right)_1 \Delta 
+ \mu H_u H_d + \mu_\Delta H_d H_d \Delta \nonumber
\end{eqnarray}

Neutrino masses come at the tree level from the terms $\mathbf{Y_{\nu, i}} L L \Delta$. It is important to note that the Yukawa couplings $\mathbf{Y_{\nu, i}}$ have a weight of 6, which is determined by the weights of $L$ (weight 3) and $\Delta$ (weight 0) involved. Here, we are exploiting the fact that for modular $\mathcal{A}_4$, at weight 6, there are two types of triplets denoted as $3a$ and $3b$ (given in App.~\ref{sec:modular}). Hence, we have two terms, $\alpha \mathbf{Y_{\nu,1}} L L \Delta$ and $\beta \mathbf{Y_{\nu,2}} L L \Delta$ for neutrino sector superpotential. In the Eq.~\eqref{eq:main_W}, charged lepton Yukawa $(\mathbf{Y_e})$ and neutrino sector Yukawas  $(\mathbf{Y_{\nu,1}}, \mathbf{Y_{\nu,2}})$ are triplets of $\mathcal{A}_4$ with weight 4 and 6 respectively, as given in Tab.~\ref{tab:Yukawa}. Note that Yukawas which have trivial transformation as modular forms are not added. The explicit form of the Yukawas in terms of Dedekind eta-function $\eta(\tau)$ of modulus $\tau$ and its derivative $\eta'(\tau)$ is provided in App.~\ref{sec:modular}. Owing to the minimal particle content of the model and the constraints imposed by the modular symmetry, the form of the leptonic mass matrices is restricted as we discuss next.

\subsection{Neutrino and Charged Lepton Mass Matrix}
For the charged lepton sector the mass matrix is given by 
\begin{equation}
    M_{\ell} =v_H
    \left(\begin{matrix}
         Y_{3,1}^{(4)}   & Y_{3,2}^{(4)} & Y_{3,3}^{(4)} \\
        Y_{3,3}^{(4)}   & Y_{3,1}^{(4)} & Y_{3,2}^{(4)} \\
        Y_{3,2}^{(4)}   & Y_{3,3}^{(4)}  & Y_{3,1}^{(4)}        
    \end{matrix}\right) \left(\begin{matrix}
        \alpha_1  & 0         & 0 \\
        0         & \alpha_2 & 0 \\
        0         & 0         & \alpha_3        
    \end{matrix}\right).
    \label{eq:mcl}
\end{equation}
On the other hand, the neutrino mass matrix is given by
\begin{equation}
    M_\nu = v_\Delta \left(\begin{matrix}
        2Y_1         &  -Y_3 & -Y_2 \\
       \ast  & 2 Y_2         & -Y_1 \\
        \ast & \ast & 2 Y_3
    \end{matrix}\right) 
    \label{eq:mnu}
\end{equation}
where $Y_i \equiv \alpha \, \pmb{Y_{3a,i}^{(6)}} + \beta \, \pmb{Y_{3b,i}^{(6)}}$ with $i \in \{1,2,3\}$, $v_{\Delta}$ is the VEV of superfield $\Delta$ and $(\ast)$ represents the symmetric part of neutrino mass matrix. Details of the diagonalization of the matrices in Eq.~\ref{eq:mcl} and \ref{eq:mnu} are shown in App.~\ref{sec:mass_mat_dia}.
The neutrino mass matrix features the interesting neutrino mass ordering independent sum rule 
\begin{equation}
    m_\text{heaviest} = \frac{1}{2} \sum_i m_i,
    \label{eq:sumrule}
\end{equation}
where $m_i$ ; $i=1,2,3$ are the three physical masses of the neutrinos and $m_\text{heaviest}$ is the heaviest out of the three light neutrinos. Note that, being a type-II seesaw mechanism, there are no heavy sterile neutrinos. This sum rule was pointed out in \cite{ChuliaCentelles:2022ogm} in an unrelated setup and can be shown by explicitly computing the invariants $\text{Tr}(m^\dagger m)$ and $\text{Tr}(m^\dagger m \, m^\dagger m)$. The explicit derivation is shown in App.~\ref{ap:sumrule}. This sum rule can be tested in different currently running and upcoming experiments. We  now explore the consequences of this sum rule in Sec.~\ref{sec:mee}.

\FloatBarrier
\section{Tests of the sum rule}
\label{sec:mee}
%
The sum rule in Eq.~\eqref{eq:sumrule} can be rewritten in an ordering-dependent way and, after imposing the mass squared differences, the neutrino masses become fixed. If we for now ignore the $< 3\%$ experimental errors in $\Delta m_{ij}^2$ we get
\begin{widetext}
\begin{align*}
    \textbf{NO:} \\
    & m_3 = m_1 + m_2 \\
    & \Delta m_{21}^2 = 7.5 \times 10^{-5} \, \text{eV}^2, \quad \Delta m_{31}^2 = 2.55 \times 10^{-3} \, \text{eV}^2 \\
    &  m_1 = 0.0282 \, \text{eV}, \quad m_2 = 0.0295 \, \text{eV}, \quad m_3 = 0.0578 \, \text{eV} \\
    \textbf{IO:} \\
    & m_2 = m_1 + m_3 \\
    & \Delta m_{21}^2 = 7.5 \times 10^{-5} \, \text{eV}^2, \quad \Delta m_{31}^2 = -2.45 \times 10^{-3} \, \text{eV}^2 \\
    & m_3 = 7.5 \times 10^{-4} \, \text{eV}, \quad m_1 = 0.049 \, \text{eV}, \quad m_2 = 0.050 \, \text{eV}
\end{align*}
\end{widetext}
where NO (IO) refers to the normal (inverted) mass ordering of the neutrinos. The $\Delta m_{ij}^2$ quoted above are the current global best fit values taken from Ref.~\cite{deSalas:2020pgw,10.5281/zenodo.4726908}.

This result has important consequences for a number of experiments. Cosmological observations are sensitive to the sum of neutrino masses which if we allow $\Delta m^2_{21}$ and $\Delta m_{31}^2$ to vary inside their $3\sigma$ regions, are predicted by our sum rule with great precision
\begin{subequations}
\label{eq:summ_nu}
\begin{align}
\sum_i m_i^\text{NO}  \in [0.1138, 0.1176] \text{ eV} \label{eq:summno}, \\ 
\sum_i m_i^\text{IO}  \in [0.1007, 0.1041] \text{ eV} \label{eq:summio}.
\end{align}
\end{subequations}
These values are compatible with the Planck 2018 results \cite{Planck:2018vyg}. Its successor, the Euclid mission, which was launched in July 2023, will probe the sum of neutrino masses with unprecedented precision, targeting a range of $\sum_i m_i = 0.03-0.06$ eV \cite{Amendola:2016saw}, and similarly the ground-based microwave background experiments CMB-S4 \cite{CMB-S4:2022ght} and SPT-3G \cite{SPT-3G:2019sok} will also be able to rule out the sum rule. As such, the sum rule's validity will be under rigorous examination in the imminent future.

\begin{widetext}
\begin{figure*}[h!t]
        \includegraphics[height=5cm]{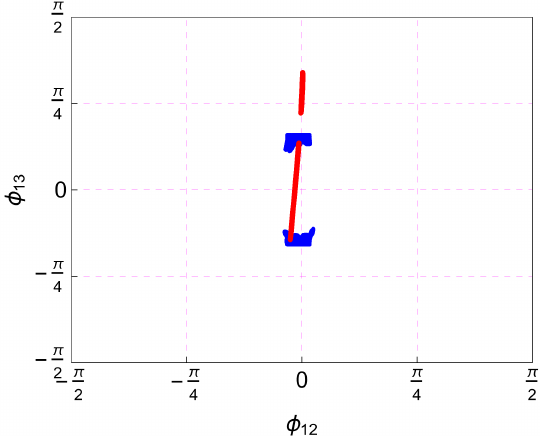} 
        \includegraphics[height=5cm]{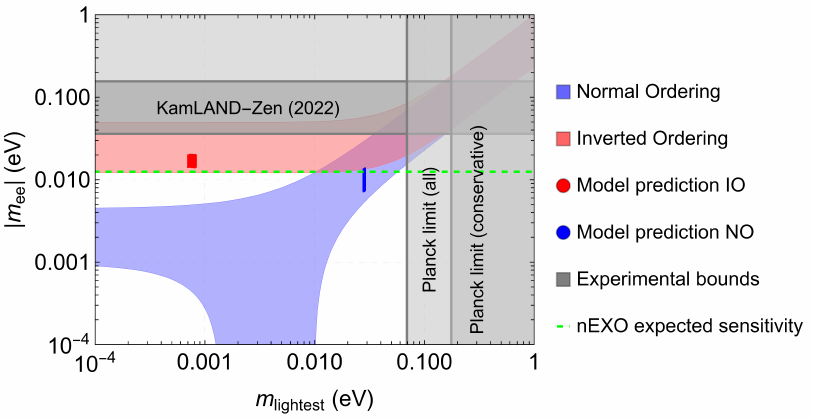}
        \caption{Neutrinoless double beta decay predictions for both NO (blue) and IO (red). \textbf{Left panel:} Correlation between Majorana phases. \textbf{Right panel:} The combination of the sum rule, the measured mass squared differences and the correlation of the Majorana phases lead to a very precise prediction for $0 \nu e e$. The KamLAND-ZEN experiment constraints the value of $|m_{ee}|$ \cite{KamLAND-Zen:2022tow} while cosmology can constraint the sum of neutrino masses \cite{Planck:2018vyg} and thus, in conjunction with the neutrino oscillation results, the mass of the lightest neutrino. The Planck limits shown are for the NO case.}
\label{fig:mee} 
\end{figure*}
\end{widetext}

Another important consequence appears in neutrinoless double beta decay experiments. The main model contribution to this process comes from the exchange of light neutrinos. In this case, the total rate is proportional to $|m_{ee}|$, defined in Eq.~\ref{eq:mee} which is also tightly predicted in our setup. Not only the neutrino masses are nearly fixed, but additionally the Majorana phases are not free but highly correlated between each other and the other oscillation parameters. This is shown for both cases in Fig.~\ref{fig:mee}.

Finally, KATRIN directly measures the effective mass of the electron neutrino defined as \cite{KATRIN:2021uub}
\\
\begin{subequations}
\begin{eqnarray}
    m_{\nu_e}^{\text{eff}} & = & \sqrt{\sum_i |U_{ei}|^2 m_i^2 }, \\
    m_{\nu_e}^{\text{eff}} & < & 0.8 \, \text{eV}, \, \text{90\% CL (2023)}, \\
    m_{\nu_e}^{\text{eff}} & < & 0.2 \, \text{eV}, \, \text{90\% CL (2025 expected sensitivity)}
\end{eqnarray}
\end{subequations}
If the mixing parameters are taken to their best fit values, the sum rule predicts
\begin{subequations}
\begin{align}
  \textbf{NO: }  m_{\nu_e}^{\text{eff}} = 0.028 \text{ eV}, \\
    \textbf{IO: } m_{\nu_e}^{\text{eff}} = 0.049\text{ eV}.
\end{align}
\end{subequations}
Therefore, if KATRIN measures the neutrino mass during its current run the model would be ruled out.
%
\section{Neutrino oscillations predictions}
\label{sec:oscillations}

Apart from the sum rule predictions, the model can also fit the mixing parameters in spite of the limited number of parameters that form the neutrino and charged lepton matrices of Eqs.~\eqref{eq:mcl} and \eqref{eq:mnu}. Before delving into the predictions of the mixing parameters let us do a parameter count. In principle the system depends on $6$ complex parameters: the modulus $\tau$, the neutrino sector free parameters $\alpha$ and $\beta$ and the $3$ charged lepton sector free parameters $\alpha_i$. Additionally, the neutrino mass scale is given by the VEV of the triplet $v_\Delta$ as in the type-II seesaw model while the charged lepton mass scale is given by the VEV of the Higgs doublet, like in the MSSM. However, without loss of generality, some considerations which will simplify the computation are in order.
\begin{itemize}
    \item In the neutrino sector we can factor out $\beta$. Then the global factor will instead be $v_\Delta \beta$ and its phase will be unphysical. Inside the neutrino matrix we are left with a dependence on $\alpha/\beta$ which we parameterize as $\alpha/\beta = r \, \text{Exp}(i \Theta)$.
    \item In the charged lepton sector we can rotate the unphysical phases of the $\alpha_i$ by redefining the right-handed fields. Therefore we can take the $\alpha_i$ to be real without loss of generality. 
    \item For a given value of $\tau$ we can solve the values of $\alpha_i$ that will lead to the correct charged lepton masses. In order to do so we can solve the invariant equations in Eq.~\eqref{eq:mclinvariants}. This leads to $6$ different solutions in the $\alpha_i$.
\end{itemize}
\begin{subequations}
\label{eq:mclinvariants}
\begin{align}
     \left|\text{Det}(M_\ell)\right| &= m_\tau m_\mu m_e\\
     \text{Tr}(M_\ell^\dagger M_\ell ) &= m_\tau^2+ m_\mu^2+ m_e^2 \\
     \text{Tr}(M_\ell^\dagger M_\ell \, M_\ell^\dagger M_\ell ) &= m_\tau^4+ m_\mu^4+ m_e^4 
\end{align}
\end{subequations}
where $m_i$; $i = e,\mu,\tau$ are the physical masses of the charged leptons.
\begin{figure}[t!]
        \includegraphics[width=10cm]{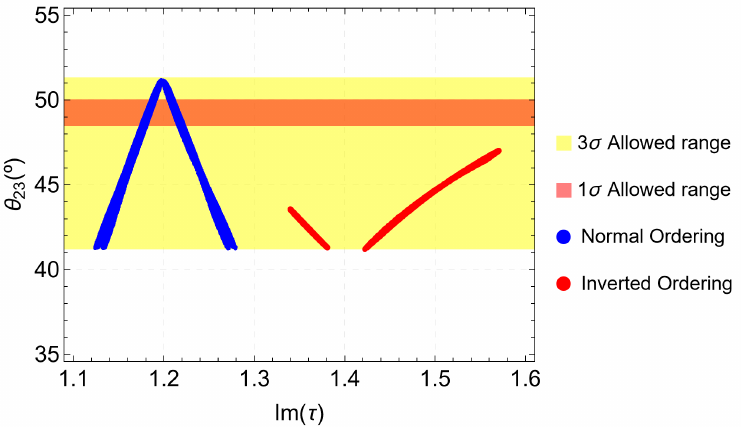} 
        \caption{The model predicts a tight correlation between the imaginary part of the modulus $\tau$ and the atmospheric angle $\theta_{23}$ in both orderings. The experimental bands correspond to the NO case and are very similar for the IO case. The model points (blue and red for NO and IO, respectively) satisfy $\theta_{ij} \in 3\sigma$ and $\delta_{CP} \in 5\sigma$. }
\label{fig:imtaucorrelations} 
\end{figure}
Therefore, the neutrino sector is determined by only $5$ real parameters $(\text{Re}(\tau),\text{Im}(\tau),r,\Theta,v_\Delta \beta)$ with the charged lepton masses already fixed to their observed values. In exchange we will obtain predictions for $9$ fundamental parameters: $3$ mixing angles $\theta_{ij}$, $3$ CP violating phases $\phi_{ij}$ and $3$ neutrino masses. Alternatively, we can rearrange these parameters into $8$ directly measurable observables: $\theta_{ij}$, $\delta_{CP}$, $|m_{ee}|$, $\Delta m^2_{ij}$ and $\sum_i m_i$. On top of the naive parameter counting the model automatically features the neutrino mass sum rule shown in Sec.~\ref{sec:mee}, which in turn fixes the absolute neutrino mass scale. It is therefore a very predictive setup as we will show explicitly in Secs.~\ref{sec:NOresults}, \ref{sec:IOresults}.

Before proceeding further let us also point out the importance of the modular symmetry in constraining the mixing angles. Since the Yukawas now transform as modular forms, their values are controlled by the $\tau$ parameter. As a result, the atmospheric angle is tightly correlated with the imaginary part of the modulus $\tau$, see Fig.~\ref{fig:imtaucorrelations}.

We now proceed to flesh out the results for the Normal and Inverted Ordering (NO and IO) of neutrino masses. As of the current date, both options are experimentally open, but the JUNO experiment is expected to begin collecting data soon, and is projected to resolve the hierarchy to the $3-4\sigma$ level over a 6-year period \cite{JUNO:2021vlw}. In what follows, we will use the results of the AHEP global fit \cite{deSalas:2020pgw,10.5281/zenodo.4726908} by imposing the 2D $1\sigma$ or $3\sigma$ constraints for the mixing angles $\theta_{ij}$. Let us also point out that the measurement of the CP violating phase $\delta_{CP}$ is not as robust as the mixing angles one, as reflected by the slight tension between Nova \cite{NOvA:2021nfi} and T2K \cite{T2K:2023smv}. For that reason, in order to reflect the lack of consensus in the measurements of $\delta_{CP}$, in our analysis we will allow it to be in its $5\sigma$ or $3\sigma$ allowed ranges.
\subsection{NO}
\label{sec:NOresults}
%
In the normal ordering of neutrino masses we have $m_3 > m_2 > m_1$, as well as $\Delta m_{31}^2 > 0$. We impose all the mixing angles and mass squared differences to be inside their 2D $3 \sigma$ regions, but for the sake of representation we let $\delta_{CP}$ in its $5\sigma$ regions. It is important to note that it is possible to obtain the mass squared differences and mixing angles inside their respective $1\sigma$ regions, while at the same time $\delta_{CP}$ can be in the $2$ or $3 \sigma$ region. This behaviour can be seen in Fig.~\ref{fig:NOcorrelations}.
The main neutrino oscillation predictions of the model in the NO case are a)  $\theta_{13} > 8.36^\circ$, which will be put to test in the next run of T2K \cite{T2K:2023smv} b) a correlation between $\theta_{13}$ and $\theta_{23}$, which may be resolved by a combination of T2K and Dune \cite{T2K:2023smv, DUNE:2020ypp} and c) A correlation between $\theta_{23}$ and $\delta_{CP}$ which may be falsified by Dune~\cite{DUNE:2020ypp}.
\vspace{2cm}
\begin{widetext}
\begin{figure*}[ht]
        \includegraphics[height=6.2cm]{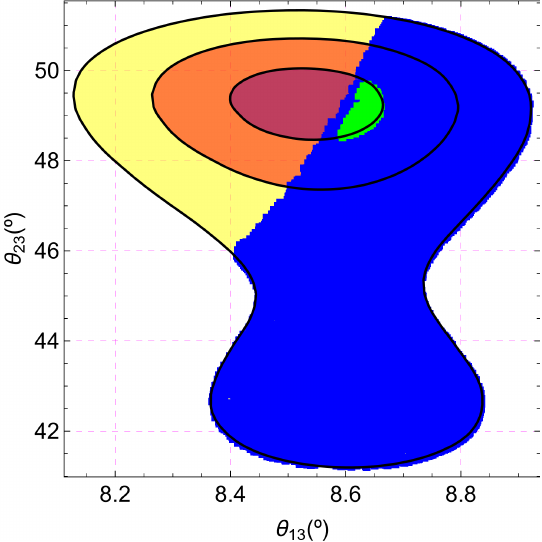}
        \hspace{0.5cm}
        \includegraphics[height=6.2cm]{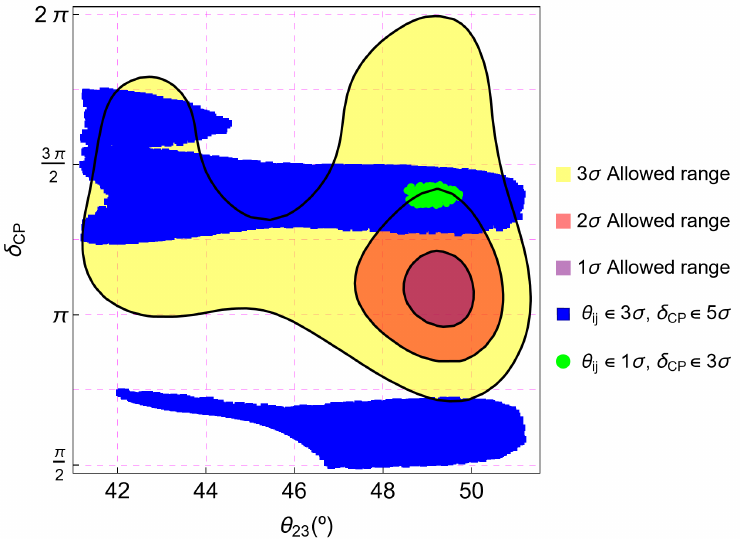}
        \caption{Model predictions in the NO case. In both panels, the blue dots satisfy all the 2D $3\sigma$ mixing angles correlations, while the CP violation phase $\delta_{CP}$ is in its $5\sigma$ region. The green dots instead have the mixing angles inside their $1\sigma$ regions and $\delta_{CP}$ in their $3\sigma$ ones. See text for details. \textbf{Left panel:} Correlation between $\theta_{13}$ and $\theta_{23}$. This correlation implies an lower bound on  $\theta_{13} > 8.36^\circ$, which may be contested in the future run of T2K \cite{T2K:2023smv}. \textbf{Right panel:} Correlation between $\theta_{23}$ and $\delta_{CP}$. This correlation may also be probed by future experiments like Dune \cite{DUNE:2020ypp}.}
\label{fig:NOcorrelations}
\end{figure*}

\begin{figure*}[ht]
        \includegraphics[height=6.2cm]{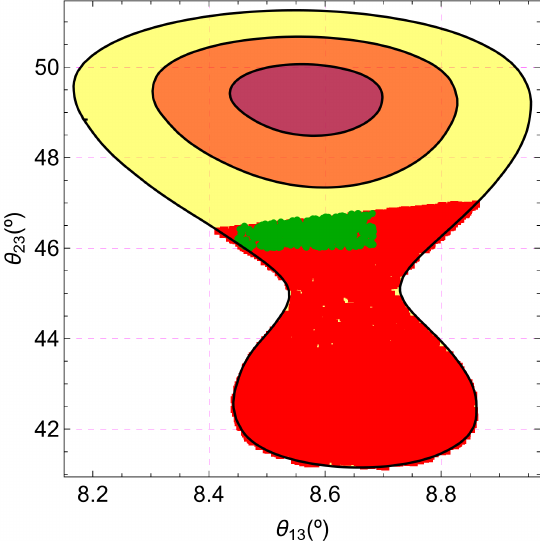}
        \hspace{0.5cm}
        \includegraphics[height=6.2cm]{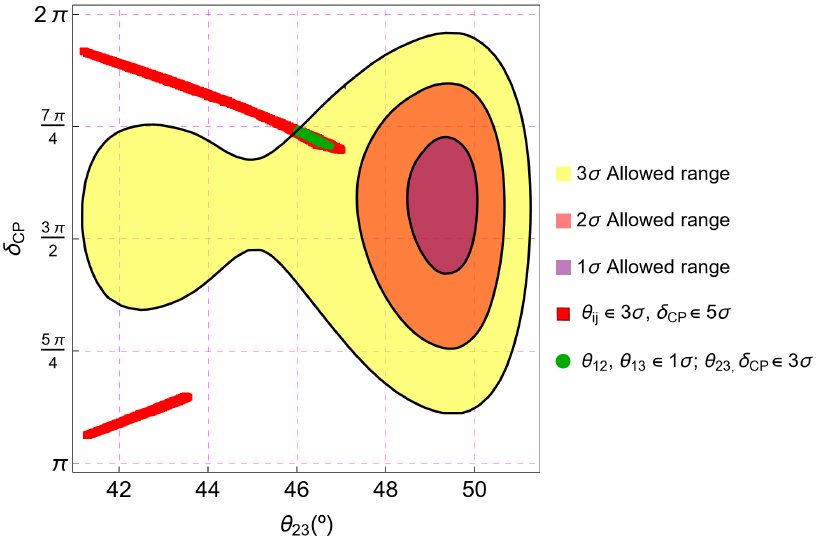}
        \caption{Model predictions in the IO case. In both panels, the red dots satisfy all the 2D $3\sigma$ mixing angles correlations, while the CP violation phase $\delta_{CP}$ is in its $5\sigma$ region. The dark green dots instead have the mixing angles $\theta_{12}$ and $\theta_{13}$ inside their $1\sigma$ regions, while $\delta_{CP}$ and $\theta_{23}$ are in their $3\sigma$ ones. See text for details. \textbf{Left panel:} $\theta_{13}$ vs $\theta_{23}$. The model predicts an upper bound on $\theta_{23} < 46.8^\circ$, which may be ruled out by future neutrino oscillation experiments. \textbf{Right panel:} Sharp correlation between $\theta_{23}$ and $\delta_{CP}$, which may be probed by future experiments like Dune \cite{DUNE:2020ypp}.}
\label{fig:IOcorrelations}
\end{figure*}
\end{widetext}
%

\subsection{IO}
\label{sec:IOresults}
Similarly, in the inverted ordering of neutrino masses, we have $m_2 > m_1 > m_3$, as well as $\Delta m_{31}^2< 0$. Unlike in the NO case, here the mixing angles $\theta_{12}$, $\theta_{13}$ and $\theta_{23}$ are nearly uncorrelated. However, $\theta_{23}$ has an upper bound of $\theta_{23} < 46.8^\circ$. Since oscillation experiments seem to prefer the upper octant, this prediction may be ruled out by Dune in the near future \cite{DUNE:2020ypp}. Moreover, $\theta_{23}$ and $\delta_{CP}$ feature a very sharp correlation, which can also be tested by Dune~\cite{DUNE:2020ypp}. This behaviour can be seen in Fig.~\ref{fig:IOcorrelations}.

The main neutrino oscillations predictions of the model in the IO case are a) $\theta_{23} < 46.8^\circ$ and b) a sharp correlation between $\theta_{23}$ and $\delta_{CP}$. If the hints that $\theta_{23}$ lie in the upper octant get confirmed, the available parameter space of this scenario will be greatly restricted.

\section{Conclusions}
\label{sec:conclusions}
We have presented a minimal extension of the MSSM based on a modular $\mathcal{A}_4$ symmetry which substantially restricts the number of parameters in the flavour space. The BSM fields and symmetries ingredients of the model are just a $SU(2)_L$ triplet $\Delta$ and modulus $\tau$, which gives rise to neutrino masses via a type-II seesaw mechanism. As a result the model is remarkably predictive. If neutrino masses follow the normal ordering, the model requires $\theta_{13} > 8.36^\circ$ as well as the correlation between $\theta_{23}$ and $\delta_{CP}$ showed in Fig.~\ref{fig:NOcorrelations}. If instead, they are arranged in the inverted ordering the predictions are $\theta_{23} < 46.8^\circ$ and an even stronger correlation between $\theta_{23}$ and $\delta_{CP}$ shown in Fig.~\ref{fig:IOcorrelations}. The combination of current and future neutrino oscillation experiments will reduce the parameter space even further and will potentially rule out the inverted ordering case.

Most importantly, the neutrino mass structure leads to a sum rule for the physical neutrino masses. Combined with neutrino oscillation data this sum rule fixes the absolute neutrino mass scale. The upcoming cosmological probes such as the Euclid mission, the CMB-S4 and SPT-3G experiments, whose first datasets are expected soon, will be able to fully test this sum rule, see Eqs.~\eqref{eq:summno} and \eqref{eq:summio}. Furthermore, the nEXO experiment will explore part of the relevant parameter space, see Fig.~\ref{fig:mee}. On the other hand, the $ m_{\nu_e}^{\text{eff}}$ value predicted by the sum rules is below KATRIN's experimental sensitivity, hence any observation in this experiment will rule out the model.

\acknowledgments

The authors would like to thank Omar Medina, Andreas Trautner and Gui-Jun Ding for helpful discussions. OP was supported in part by the National Natural Science Fund of China Grant No. 12350410373.
\appendix

\section{Diagonalization of mass matrices and parametrization of unitary matrices}
\label{sec:mass_mat_dia}

\begin{figure}[h!]
    \centering
    \includegraphics[height=4cm]{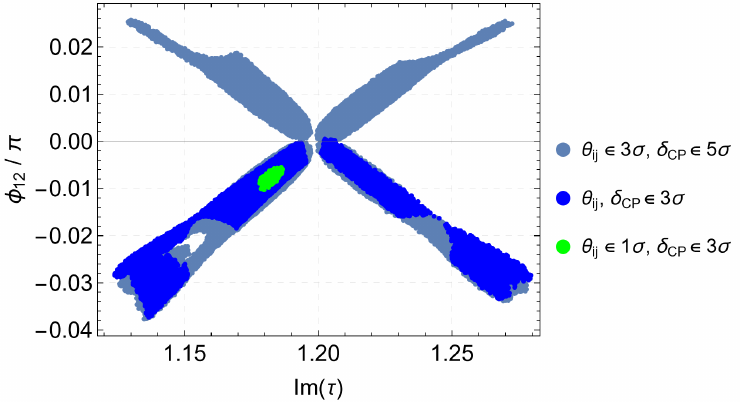}
    \includegraphics[height=4cm]{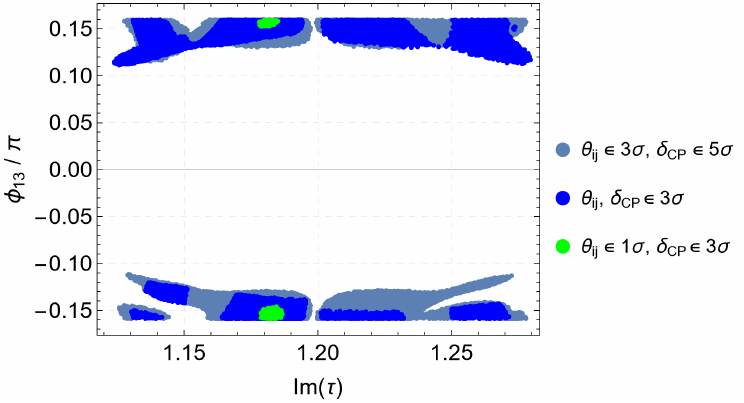}
    \caption{Correlation of Majorana phases with $\text{Im}(\tau)$ in the NO scenario.}
    \label{fig:phasesNO}
\end{figure}

\begin{figure}[h!]
    \centering
    \includegraphics[height=4cm]{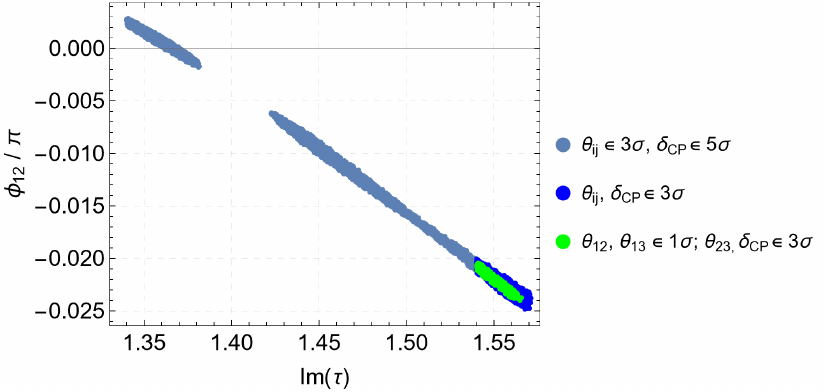}
    \includegraphics[height=4cm]{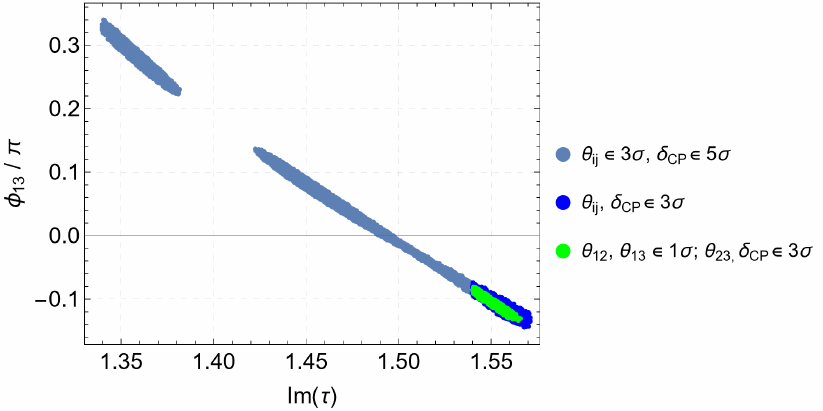}
    \caption{Correlation of Majorana phases with $\text{Im}(\tau)$ in the IO scenario.}
    \label{fig:phasesIO}
\end{figure}

The mass matrices in Eqs.~\eqref{eq:mcl} and \eqref{eq:mnu} are diagonalized as follows
\begin{align*}
    U_\nu^T M_\nu U = \text{diag}(m_1, m_2, m_3) \\
    U_\ell^\dagger M_\ell V = \text{diag}(m_e, m_\mu, m_\tau) \\
    U_\text{lep} = U_\ell^\dagger U_\nu
\end{align*}
where $U_\text{lep}$ is the lepton mixing matrix which parametrizes the interaction between the $W$ boson and the leptons and is probed by neutrino oscillation experiments. In the symmetric parametrization \cite{Schechter:1980gr, Rodejohann:2011vc} a general unitary matrix can be written as
\begin{equation*}
U_\text{lep} = P(\delta_1, \delta_2, \delta_3) \, U_{23}(\theta_{23}, \phi_{23}) \, U_{13} (\theta_{13}, \phi_{13}) \, U_{12}(\theta_{12}, \phi_{12}) \, ,
\end{equation*}
where $P(\delta_1, \delta_2, \delta_3)$ is a diagonal matrix of unphysical phases and the $U_{ij}$ are complex rotations in the $ij$ plane, as for example,
\begin{equation*}
    U_{23} (\theta_{23}, \phi_{23}) = \left(\begin{matrix}
        1 & 0 & 0 \\
        0 & \cos\theta_{23} & \sin\theta_{23}\, e^{-i \phi_{23}} \\
        0 & -\sin\theta_{23} \,e^{i \phi_{23}} & \cos\theta_{23}
        \end{matrix} \right) \,.
\end{equation*}
The phases $\phi_{12}$ and $\phi_{13}$ are relevant for $0\nu e e$ decay, while the combination $\delta_{CP} = \phi_{13} - \phi_{12} - \phi_{23}$ is the usual Dirac $CP$ phase measured in neutrino oscillations. 
The primary contribution to the $0\nu e e$ decay process is the exchange of light neutrinos. The effective mass $|m_{ee}|$, which determines the decay rate, is given in the symmetric parametrization \cite{Schechter:1980gr, Rodejohann:2011vc} as
\begin{equation}
 \label{eq:mee}
|m_{ee}| = \left|\sum_i U_{ei}^2 m_i\right| = \left|c_{12}^2 c_{13}^2 m_1 + s_{12}^2 c_{13}^2 e^{2 i \phi_{12}} m_2 + s_{13}^2 e^{2 i \phi_{13}} m_3\right|,
\end{equation}
where \(c_{ij} = \cos \theta_{ij}\), \(s_{ij} = \sin \theta_{ij}\) represent the mixing angles, \(m_i\) are the neutrino masses, and \(\phi_{12}\), \(\phi_{13}\) are the CP-violating phases.
Our model's predictions for the $0\nu e e $ decay are illustrated in the right panel of Fig.~\ref{fig:mee}  in the main text, plotting $|m_{ee}|$ against the lightest neutrino mass, $m_{\text{lightest}}$, which corresponds to $m_1$ in the NO case and to $m_3$ in the IO case. The light blue and red regions correspond to the $3\sigma$ allowed ranges for normal ordering (NO) and inverted ordering (IO) scenarios, respectively. The dark blue and red regions represent our model's predictions for the NO and IO scenarios, respectively. As depicted in Fig.~\ref{fig:mee}, our model confines the predicted $|m_{ee}|$ to a narrow range for both NO and IO scenarios, which is attributable to the specific values of the mixing angles, neutrino masses, and Majorana phases constrained by our model. In particular, the width of each model band is restricted by the combination of the sum rule and the measured mass squared differences, which sets a narrow range for $m_\text{lightest}$ in each ordering. On the other hand, the correlations between mixing parameters in the model, most importantly the Majorana phases $\phi_{12}$ and $\phi_{13}$, also limit the height of the model region, which would cover the entirety of the light colored region for a given value of $m_\text{lightest}$ if they were free inside the experimental 3$\sigma$ ranges. See Figs.~\ref{fig:phasesNO} and \ref{fig:phasesIO}, which highlight the tight relation between the Majorana phases and $\tau$ in NO and IO, respectively.

Furthermore, the gray shaded region along the horizontal axis in Fig.~\ref{fig:mee} represents the Planck collaboration's constraint on $\sum_i m_i$ translated into a constraint on $m_{\text{lightest}}$ for the NO case \cite{Planck:2018vyg}. For clarity, this constraint is not depicted for the IO scenario, but see the main text for further details. Note that the model prediction of $\sum_i m_i$ is very similar for both orderings, around $0.1$ eV. The gray shaded area on the vertical axis indicates the upper limit on $|m_{ee}|$ from the KamLAND-ZEN experiment \cite{KamLAND-Zen:2022tow}, while the dashed green line shows the future sensitivity expected from the nEXO experiment \cite{nEXO:2017nam}.

\section {$\mathcal{A}_4$ Multiplication rule and Modular Yukawa construction}
\label{sec:modular}
\textbf{$\mathcal{A}_4$ symmetry:} $\mathcal{A}_4$ is an even permutation group of four objects. It is also the symmetry group of a regular tetrahedron. It has 4!/2=12 elements and can be generated by two generators $S$ and $T$ obeying the relations:
\begin{equation*}
    S^2=T^3=(ST)^3= \mathcal{I} .
\end{equation*}
The group has four irreducible representations, a trivial singlet 1, two non-trivial singlet $1'$, $1''$, and a triplet 3. The product rules for the singlets and triplet are:
\begin{eqnarray} \label{eq:A4rule}
    &1 \otimes 1&=1=1'\otimes 1'', \quad 1'\otimes 1'=1'', \quad \quad 1'' \otimes 1''=1'.  \nonumber \\
    &3 \otimes 3&=1 \oplus 1' \oplus 1'' \oplus 3_S \oplus 3_A .
\end{eqnarray}
where, $3_{S(A)}$ denotes the symmetric (and anti-symmetric) combination. In the complex basis where $T$ is a diagonal matrix, we have,  
\begin{align*}
  S= \frac{1}{3}\begin{pmatrix}
  -1 & 2 & 2 \\
 2 & -1 &2\\
 2&2 & -1 
 \end{pmatrix}, \quad \quad
  T= \begin{pmatrix}
  1 & 0 & 0 \\
  0 & \omega & 0  \\
  0& 0& \omega^2
  \end{pmatrix}.   
\end{align*}
where $\omega$ is the cubic root of unity. Given two triplets $a=(a_1,a_2,a_3)$ and $b=(b_1,b_2,b_3)$, their product decomposes following Eq.~\eqref{eq:A4rule} and they are expressed as:
\begin{eqnarray}
    &(ab)_1 &= a_1b_1 + a_2b_3 + a_3b_2 \nonumber \\
    &(ab)_{1'} &= a_3b_3 + a_1b_2 + a_2b_1 \nonumber \\
    &(ab)_{1''} &= a_2b_2 + a_3b_1 + a_1b_3 \nonumber \\
     &(ab)_{3_S} &=\frac{1}{\sqrt{3}}\begin{pmatrix}
  2a_1b_1 - a_2b_3 - a_3b_2 \\
 2a_3b_3 - a_1b_2 - a_2b_1  \\
 2a_2b_2 - a_3b_1 - a_1b_3  
 \end{pmatrix} \nonumber \\
     &(ab)_{3_A} &=\begin{pmatrix}
 a_2b_3 - a_3b_2 \\
  a_1b_2 - a_2b_1  \\
  a_3b_1 - a_1b_3  
 \end{pmatrix}
\end{eqnarray}
\par \textbf{Modular framework:} Here we summarize modular symmetry framework in context of $\mathcal{A}_4$ symmetry. The modular group $\bar{\Gamma}$ is the group of  linear fractional transformations $\gamma$ which acts on the complex variable  $\tau$ linked to the upper-half complex plane as follows:
\begin{equation}\label{eq:tau-SL2Z}
\tau \longrightarrow \gamma\tau= \frac{a\tau + b}{c \tau + d}
\end{equation}
${\rm where,}~~ a,b,c,d \in \mathbb{Z}~~ {\rm and }~~ ad-bc=1, 
~~ {\rm Im} [\tau]>0 $.
The modular group is  isomorphic to the transformation $PSL(2,\mathbb{Z})=SL(2,\mathbb{Z})/\{\mathcal{I},-\mathcal{I}\}$, and it is generated by two elements $S$ and $T$ satisfying
\begin{equation*}
S^2 =\mathcal{I}\ , \qquad (ST)^3 =\mathcal{I}\ .
\end{equation*}
Representing $S$ and $T$ as 
\begin{align*}
     S= \begin{pmatrix}
  0 & 1  \\
 -1 & 0
 \end{pmatrix}, \quad \quad
  T= \begin{pmatrix}
  1 & 1  \\
  0 & 1  
  \end{pmatrix}, 
\end{align*}
then they correspond to the following transformations,
\begin{eqnarray*}
S:\tau \longrightarrow -\frac{1}{\tau}\ , \quad \quad
T:\tau \longrightarrow \tau + 1\ .
\end{eqnarray*}
The group $S L (2, \mathbb{Z})=\Gamma(1) \equiv \Gamma $ contains a series of infinite normal subgroups $\Gamma(N),~ (N=1,2,3,\dots)$ and defined as
 \begin{align*}
 \Gamma(N)= \left \{ 
 \begin{pmatrix}
 a & b  \\
 c & d  
 \end{pmatrix} \in SL(2,\mathbb{Z})~ ,
 \begin{pmatrix}
  a & b  \\
 c & d  
 \end{pmatrix} =
  \begin{pmatrix}
  1 & 0  \\
  0 & 1  
  \end{pmatrix} ({\rm mod} N) \right \}
\end{align*}
Definition of $\bar\Gamma(2)\equiv \Gamma(2)/\{\mathcal{I},-\mathcal{I}\}$ for $N=2$. Since $-\mathcal{I}$ is not associated with $\Gamma(N)$ for $N>2$ case, one can have $\bar\Gamma(N)= \Gamma(N)$, which are infinite normal subgroups of $\bar \Gamma$ known as principal congruence subgroups. The quotient groups  $\Gamma_N\equiv \bar \Gamma/\bar \Gamma(N)$  are called finite modular groups. The condition of $T^N=\mathcal{I}$ is applied to these finite groups $\Gamma_N$. For small $N$ ($\leq 5$), the groups $\Gamma_N$  are isomorphic to permutation groups~\cite{deAdelhartToorop:2011re}. Namely,  $\Gamma_2\simeq S_3$, $\Gamma_3\simeq A_4$, $\Gamma_4\simeq S_4$ and $\Gamma_5\simeq A_5$. 

Modular forms $f(\tau)$ of weight $\mathit{k}$ and level $N$ are holomorphic functions of the complex variable $\tau$ and its transformation under the group $\Gamma(N)$ is given as follows:
\begin{equation*}
f(\gamma\tau)= (c\tau+d)^{\mathit{k}} f(\tau)~, ~~ \gamma=\begin{pmatrix}
  a & b  \\
  c & d  
  \end{pmatrix} \in \Gamma(N)~ ,
\end{equation*}
where $\mathit{k}$ is even and non-negative. Modular forms of weight $\mathit{k}$ and level $N$ constitute a finite-dimensional linear space. Within this space, it is possible to find a basis where a multiplet of modular forms $f_i(\tau)$ undergoes transformations following a unitary representation $\rho$ of the finite group $\Gamma_N$:
\begin{equation*}
f_i(\gamma\tau)= (c\tau+d)^{\mathit{k}} \rho_{ij}(\gamma)f_j(\tau)~, ~~ \gamma \in \Gamma(N)~ ,
\end{equation*}
A field $\phi^{(I)}$ transforms as given in Eq.~\eqref{eq:field-SL2Z}.
%
 The scalar fields kinetic terms are given as follows
\begin{equation*}
\sum_I\frac{|\partial_\mu\phi^{(I)}|^2}{(-i\tau+i\bar{\tau})^{\mathit{k}_I}} ~,
\label{kinetic}
\end{equation*}
which doesn't change under the modular transformation, and eventually, the overall factor is absorbed by the field redefinition.

Thus, the Lagrangian should be invariant under the modular symmetry.
Our model is based on  $\mathcal{A}_4$ ($N=3$) modular group. The modular forms of the Yukawa coupling  $\mathbf{Y^{(2)}}$ = {$(y_1,y_2,y_3)$}  with weight 2, which transforms
as a triplet of $\mathcal{A}_4$ can be expressed in terms of Dedekind eta-function  $\eta(\tau)$ and its derivative \cite{Feruglio:2017spp}:
\begin{eqnarray} 
\label{eq:Y-A4}
y_1(\tau) &=& \frac{i}{2\pi}\left( \frac{\eta'(\frac{\tau}{3})}{\eta(\frac{\tau}{3})}  +\frac{\eta'(\frac{\tau +1}{3})}{\eta(\frac{\tau+1}{3})}  
+\frac{\eta'(\frac{\tau +2}{3})}{\eta(\frac{\tau+2}{3})} - \frac{27\eta'(3\tau)}{\eta(3\tau)}  \right), \nonumber \\
y_2(\tau) &=& \frac{-i}{\pi}\left( \frac{\eta'(\frac{\tau}{3})}{\eta(\frac{\tau}{3})}  +\omega^2\frac{\eta'(\frac{\tau +1}{3})}{\eta(\frac{\tau+1}{3})}  
+\omega \frac{\eta'(\frac{\tau +2}{3})}{\eta(\frac{\tau+2}{3})}  \right) , \label{eq:Yi} \\ 
y_3(\tau) &=& \frac{-i}{\pi}\left( \frac{\eta'(\frac{\tau}{3})}{\eta(\frac{\tau}{3})}  +\omega\frac{\eta'(\frac{\tau +1}{3})}{\eta(\frac{\tau+1}{3})}  
+\omega^2 \frac{\eta'(\frac{\tau +2}{3})}{\eta(\frac{\tau+2}{3})}  \right)\,.
\nonumber
\end{eqnarray}
The  expression of Dedekind eta-function  $\eta(\tau)$ is given by:
\begin{equation*}
    \eta(\tau)=q^{1/24} \prod^{\infty}_{n=1}\left( 1-q^n \right), \quad \quad q \equiv e^{i 2 \pi \tau}. 
\end{equation*}
In the form of q-expansion, the modular Yukawa of Eq.~\eqref{eq:Y-A4} can be expressed as:
\begin{eqnarray*}
   y_1(\tau) &=& 1+12q +36q^2 + 12q^3+... \\
 y_2(\tau) &=& -6q^{1/3}(1+7q+8q^2+...) \\
   y_3(\tau) &=& -18q^{2/3}(1+2q+5q^2+...) \quad .
\end{eqnarray*}
From the q-expansion we have the following constraint for modular Yukawa couplings:
\begin{eqnarray} \label{eq:modularconstr}
    y_2^2+2y_1y_3=0.
\end{eqnarray}
Higher modular weight Yukawa couplings can be constructed from weight 2 Yukawa ($\mathbf{Y^{(2)}}$) using the $\mathcal{A}_4$ multiplication rule. For modular weight $\mathit{k}=4$, we have the following Yukawa couplings: 
\begin{eqnarray*}
    &Y^{(4)}_3 &= (y_1^2-y_2y_3,y_3^2-y_1y_2,y_2^2-y_1y_3) \\
     &Y^{(4)}_1 &= y_1^2 + 2y_2 y_3 \\
    &Y^{(4)}_{1'} &= y_3^2 + 2y_1 y_2 \\
    &Y^{(4)}_{1''} &= y_2^2 + 2y_1 y_3 \\
\end{eqnarray*}
At modular weight  $\mathit{k}=6$, the Yukawa couplings are:
\begin{eqnarray*} 
    &Y^{(6)}_1 &=y_1^3+y_2^3+y_3^3-3y_1 y_2 y_3 \nonumber \\
    &Y^{(6)}_{3a} &=(y_1^3+2y_1y_2y_3, y_1^2y_2+2y_2^2y_3, y_1^2y_3+2y_3^2y_2)  \\
    &Y^{(6)}_{3b} &=(y_3^3+2y_1y_2y_3, y_3^2y_1+2y_1^2y_2, y_3^2y_2+2y_2^2y_1)  \\
    &Y^{(6)}_{3c} &=(y_2^3+2y_1y_2y_3, y_2^2y_3+2y_3^2y_1, y_2^2y_1+2y_1^2y_3) 
\end{eqnarray*}
Due to the constraint mentioned in Eq.~\eqref{eq:modularconstr}, we see that $Y^{(4)}_{1''}=0$ and $Y^{(6)}_{3c}=0$. In general, the  Dimensions ($d_\mathit{k}$) of modular forms of the level 3 and weight $\mathit{k}$ is $\mathit{k}+1$~\cite{Feruglio:2017spp,Kikuchi:2022txy}. The representations for different weights are shown in Tab.~\ref{tab:represntations}.
\begin{table}[ht]
    \centering
    \begin{tabular}{c|c|c}
        \hline \hline
Weight $(\mathit{k})$ & $d_\mathit{k}$ & $\mathcal{A}_4$ representations \\ 
\hline
 2 & 3 & $\pmb{3}$ \\ 
4 & 5 & $\pmb{3}$+$\pmb{1}$+$\pmb{1^{\prime}}$ \\ 
6 & 7 & $\pmb{3}$+$\pmb{3}$+$\pmb{1}$ \\  
8 & 9 & $\pmb{3}$+$\pmb{3}$+$\pmb{1}$+$\pmb{1^{\prime}}$+$\pmb{1^{\prime \prime}}$ \\ 
10 & 11 & $\pmb{3}$+$\pmb{3}$+$\pmb{3}$+$\pmb{1}$ +$\pmb{1^{\prime}}$ \\ 
   \hline 
    \end{tabular}
    \caption{$\mathcal{A}_4$ representations for different weight $\mathit{k}$.}
    \label{tab:represntations}
\end{table}
\section{Proof of sum rule}
\label{ap:sumrule}

We can write the neutrino mass invariants in terms of $r^2 \equiv |Y_1|^2 +|Y_2|^2 +|Y_3|^2$
%
\begin{align*}
    \text{Tr}(M_{\nu}^\dagger M_{\nu}) = 6 r^2 \\
    \text{Tr}(M_{\nu}^\dagger M_{\nu} \, M_{\nu}^\dagger M_{\nu}) = 18 r^4 
\end{align*}
Therefore, we find $\frac{1}{2} \text{Tr}(M_{\nu}^\dagger M_{\nu}) ^2 =  \text{Tr}(M_{\nu}^\dagger M_{\nu} \, M_{\nu}^\dagger M_{\nu}) $, which in turn implies
\begin{align*}
\frac{1}{2} (m_1^2 + m_2^2 + m_3^2)^2 = m_1^4 + m_2^4 + m_3^4
\end{align*}
We can solve one of the masses, for example for $m_3$
\begin{equation}
m_3^2 = (m_1 \pm m_2)^2    
\end{equation}
And after imposing that the masses are positive we find a unique solution for each ordering
\begin{align*} 
m_3 ^{NO} = m_1^{NO} + m_2^{NO}, \hspace{0.5cm} m_3^{NO} > m_2^{NO}> m_1^{NO}\\
m_2 ^{IO} = m_1^{IO} + m_3^{IO}, \hspace{0.5cm} m_2^{IO} > m_1^{IO}> m_3^{IO}
\end{align*}
Or alternatively, adding the heaviest of the masses in both sides
\begin{equation}
    m_\text{heaviest} = \frac{1}{2} \sum_i m_i
\end{equation}
Since we know the mass squared difference of neutrino masses this sum rule leads to a definite prediction of the neutrino mass scale.
\section{$\Delta$ Scalar Mass and $\mu_\Delta$}
\label{sec:scalar_mass}
Starting with Eq.~\eqref{eq:main_W} the scalar potential can be obtained from 
\begin{align}
    V &= V_F + V_D \nonumber \\
    &= F^{*i} F_{i} + \frac{1}{2} \displaystyle\sum_a D^a D^a \nonumber \\
    &= W^*_i W^i + \frac{1}{2} \displaystyle\sum_a g_a^2 \left(\phi^\dagger T^a \phi\right)^2,
\end{align}
where $V_F$ and $V_D$ stand for F and D potential contributions, respectively. $W^i=\frac{\delta W}{\delta \phi_i}$, $\phi$ stands for superfields from Tab.~\ref{tab:particles} of the main text, $g_a$'s are the gauge coupling constants, and $T^a$'s are the generators of the gauge symmetries. 

For the sake of simplicity, to obtain the $\Delta$ scalar and Higgs masses we assume $\mathcal{R}-$parity conservation and focus on $\mathcal{R}-$parity even, electrically neutral, and real scalars. The corresponding mass matrix elements are given as

\begin{subequations}
    \label{eq:s_mass}
    \begin{align}
        m_{h_u h_u}^2 &= \mu^2 + \left( 2 g_2^2 + g_1^2\right) \frac{3}{8} v_u^2, \\
        m_{h_d h_d}^2 &= \mu^2 + \left( 2 \mu_\Delta^2 + \frac{g_2^2}{4} + \frac{g_1^2}{8}\right) 3 v_d^2 + 2 \mu_\Delta^2 v_\Delta^2, \\ 
        m_{\Delta \Delta}^2 &= \left(2 g_1^2 + \frac{g_2^2}{2}\right) 3 v_\Delta^2 + 2 \mu_\Delta^2 v_d^2, \\
        m_{h_u h_d}^2 &= \mu \mu_\Delta \frac{v_\Delta}{\sqrt{2}}, \\
        m_{h_u \Delta}^2 &= \mu \mu_\Delta \frac{v_d}{\sqrt{2}}, \\
        m_{h_d \Delta}^2 &= 2 \mu_\Delta^2 v_d v_\Delta + \mu \mu_\Delta \frac{v_u}{\sqrt{2}},
    \end{align}
\end{subequations}
in the $(h_u^0, h_d^0, \Delta^0)$ basis. In the limit $\mu_\Delta\rightarrow 0$ we recover MSSM scenario, where $v_\Delta\rightarrow0$, and all mixing terms in Eq.~\eqref{eq:s_mass} become zero. Hence, if $\mu_\Delta\rightarrow0$ $H_u$ and $H_d$ generate masses for up and down quark and leton sectors, respectively. Furthermore,
\begin{subequations}
    \label{eq:MSSM_defs}
    \begin{align}
        &v_u^2 + v_d^2 = v^2 = (246~\text{GeV})^2, \\
        &\frac{v_u}{v_d} = \tan (\beta),
    \end{align}
\end{subequations}
are defined as usual. The potential minimization equations are given as
\begin{subequations}
    \label{eq:v_min_eq}
    \begin{align}
        \frac{\partial V}{\partial h_u} &= 0 \rightarrow \nonumber \\
        &\mu^2 + \left( 2 g_2^2 + g_1^2 \right) \frac{v_u^2}{8} + \sqrt{2} \mu \mu_\Delta \frac{v_\Delta v_d}{v_u} = 0, \\
        \frac{\partial V}{\partial h_d} &= 0 \rightarrow \nonumber \\
        &\mu^2 + \left( 16 \mu_\Delta^2 + 2 g_2^2 + g_1^2 \right) \frac{v_d^2}{8} \nonumber \\
        &+ 2 \mu_\Delta^2 v_\Delta^2 + \sqrt{2} \mu \mu_\Delta \frac{v_\Delta v_u}{v_d} = 0, \\
        \frac{\partial V}{\partial \Delta^0_R} &= 0 \rightarrow \nonumber \\
        &\left( g_2^2 + 4 g_1^2 \right) \frac{v_\Delta^2}{2} + 2 \mu_\Delta^2 v_d^2 + \sqrt{2} \mu \mu_\Delta \frac{v_d v_u}{v_\Delta} = 0.
    \end{align}
\end{subequations}

Taking $\mu_\Delta \ll 1$, due to seesaw-II scenario, $v_\Delta \ll v_{u,d}$ is obtained via $v_\Delta^3 \approx - \sqrt{2} \mu \mu_\Delta v_u v_d / (2 g_1^2 + g_2^2/2)$. To satisfy the $\Delta$ scalar mass experimental constraint, the SUSY soft breaking term for $\Delta$ scalar must dominate the mass contribution to it, $m_\Delta^2 \approx m_{\Delta, soft}^2 + \mathcal{O} (1~\text{GeV})^2 \approx \mathcal{O} (1~\text{TeV})^2$. As a result, the $\Delta$ scalar mass dominantly depends on the value of the SUSY soft breaking term $m_{\Delta, soft}^2$, assuming the case of $\mu_\Delta \ll 1$ for seesaw-II purposes. Finally, the mass square of $\Delta$ under aforementioned conditions, $v_\Delta^3 \approx - \sqrt{2} \mu \mu_\Delta v_u v_d / (2 g_1^2 + g_2^2/2)$ approximation, as well as taking into account that $g_1^2 (m_W) = 0.2136$, $g_2^2 (m_W) = 0.4210$, and using Eq.~\eqref{eq:s_mass} (including the soft SUSY breaking term) is given as
\begin{align}
    m_\Delta^2 &\approx m_{\Delta, soft}^2 + \left( 2 g_1^2 + \frac{g_2^2}{2} \right) 3 v_\Delta^2 + 2 \mu_\Delta^2 v_d^2 \nonumber \\
    &\approx m_{\Delta, soft}^2 + 2 v_\Delta^2 + 2 \mu_\Delta^2 v_d^2.
\end{align}


%

%
\bibliographystyle{utphys}
\providecommand{\href}[2]{#2}\begingroup\raggedright\endgroup
\end{document}